\documentclass{aa} 
\usepackage{graphicx}
\begin{document} 
 
\title{Probing the structure of a birthplace of intermediate-mass stars:
Ammonia cores in Lynds~1340} 
 
\author{M. Kun\inst{1} \and J.G.A. Wouterloot\inst{2} \and L.V. T\'oth\inst{3}\thanks{\emph{Present address:}
Max-Planck-Institut f\"ur Astronomie, K\"onigstuhl~17, D-69117~Heidelberg, Germany}} 
\institute{Konkoly Observatory, H-1525 Budapest, P.O.Box 67, Hungary 
\and Joint Astronomy Centre, 660 N. A'ohoku Place, University Park, 96720 Hilo,
Hawaii, USA 
\and Department of Astronomy, Lor\'and E\"otv\"os University, H-1117 Budapest, 
P\'azm\'any P\'eter s\'et\'any 1, Hungary} 
\offprints{M. Kun
\email{kun@konkoly.hu}} 
\date{Received  / Accepted  }
  
\abstract{
Lynds~1340, a molecular cloud forming intermediate-mass stars, has been mapped in the 
NH$_3$(1,1) and (2,2) transitions with the Effelsberg 100-m telescope. 
We observed the whole area of the cloud where C$^{18}$O 
emission was detected earlier, at a 40$\arcsec$ grid, 
with additional positions towards the C$^{18}$O peaks 
and optically invisible  IRAS  point sources. 
Our observations covered an area of 170~arcmin$^2$, corresponding 
to about 5.15~pc$^2$ at a distance of 600~pc, and revealed 10 ammonia 
cores. The cores, occupying some 7\,\% of the mapped area, probably represent the 
highest density regions of L\,1340. Their total mass is $\sim$80\,M$_{\sun}$, 
about 6\,\% of the mass traced by C$^{18}$O. Six cores are associated with 
optically invisible IRAS point sources. Their average nonthermal line 
width is 0.78\,km\,s$^{-1}$, while the same quantity for the four 
starless cores is 0.28\,km\,s$^{-1}$. We suggest that the 
narrow-line cores are destined to form low-mass stars, whereas small groups of 
intermediate-mass stars are being formed in the turbulent cores.
The features traced by NH$_3$, $^{13}$CO, 
C$^{18}$O and \ion{H}{i} obey the line width--size 
relation $\Delta v_\mathrm{NT} \propto R_{1/2}^{0.41}$.
Comparison of sizes, densities and  nonthermal line widths of ammonia cores 
with those of C$^{18}$O and $^{13}$CO structures supports
the scenario in which core formation has been induced by turbulent 
fragmentation.
The typical physical properties of the NH$_3$ cores of L\,1340,
$\langle R_{1/2} \rangle$=0.08\,pc, $\langle T_\mathrm{kin} \rangle$=13.8\,K, 
$\langle \Delta v_\mathrm{total} \rangle$=0.64\,km\,s$^{-1}$, and 
$\langle M \rangle$=9\,M$_{\sun}$ are 
close to those of the high-mass star forming Perseus and Orion~B clouds.
}
\titlerunning{Ammonia in L\,1340}
\authorrunning{Kun et al.}
\maketitle

\section{Introduction} 
\label{Intr} 
 
Observations of nearby star forming regions have shown that high 
mass stars are born as members of dense clusters in massive 
molecular cloud cores, whereas small, cold  cores give 
birth to one or a few solar type stars. The transition from 
isolated to clustered mode of star formation occurs in molecular 
clouds forming intermediate mass stars (Testi, Palla \& Natta~\cite{TPN}). 
Molecular cores forming isolated low-mass stars and rich clusters differ 
from each other in structure, physical properties and evolution 
(e.g. Caselli \& Myers~\cite{CM}). Observations of various 
star forming regions are helpful in getting an unbiased view on the
physical processes leading to a particular mode of star formation. 
We studied the structure of Lynds~1340, a medium-mass molecular cloud   
in order to compare the properties of this HAe/Be star birthplace with 
well-studied star forming environments. 

The star-forming molecular cloud Lynds~1340 is located in Cassiopeia, 
near (l,b)=(130,11), at a distance of 600\,pc from the Sun (Kun et al.~\cite{Paper1}, 
hereafter Paper~I). In optical images, the cloud is visible as 
the faint, blue reflection nebula DG~9 
(Dorschner \& G\"urtler~\cite{DG}),  which is
illuminated by $B$ and $A$ type stars. RNO~7, 8 and 9 (Cohen~\cite{RNO})
associated with the cloud are probably signposts of recent star 
formation in L\,1340. 
 
Paper~I presented $^{13}$CO and C$^{18}$O maps of L\,1340, 
obtained with the 4-m radio telescope of Nagoya University, 
its distance determination, and a list of candidate young stellar objects.
The $^{13}$CO observations revealed a molecular mass of 1200\,M$_{\sun}$,
distributed in three clumps denoted as {\em cores\/} {\it A\/}, {\it B\/} 
and {\it C\/}, each associated with a number of  IRAS  point 
sources and H$\alpha$ emission stars. Following the definition given 
by Blitz \& Williams~(\cite{BW}), however, it is more appropriate to refer 
to these $^{13}$CO substructures as {\em clumps\/}, which  
may contain higher density cores. 
 
Yonekura et al.'s~(\cite{YDMOF}) $^{13}$CO survey results have shown
this cloud to be fairly isolated, instead of being
a part of a giant molecular complex. They also pointed out 
that earlier Nagoya $^{13}$CO and C$^{18}$O measurements calibrated 
with S140, including those presented in Paper~I, should be revised due 
to the incorrect value of 6\,K used for the $^{13}$CO 
radiation temperature of S140. (C$^{18}$O measurements were 
calibrated assuming the same ratio for $T_{R}^{*}/T_A$ as that for the
$^{13}$CO line.) Using the revised value of 9\,K, column densities and 
volume densities derived from the optically thin C$^{18}$O
line can easily be corrected.

Recently, Kumar, Anandarao \& Yu~(\cite{KAY}) found three optical 
HH objects, HH~487, 488, and 489, emerging from YSOs in L\,1340\,A.
The same work also has shown that RNO\,7 is a compact cluster 
of some 26 stars, and predicted a mass about  6\,M$_{\sun}$ for 
its brightest member.  Recent spectroscopic study of the 
illuminating stars of DG\,9 and the candidate YSOs listed 
in Paper~I suggests that L\,1340 is a birthplace of small groups 
of low and intermediate mass stars (Kun~\cite{K2002a}). No O-type 
stars have been formed in this cloud. Both its size 
and star forming properties represent an intermediate mode between the 
isolated low mass star formation and clustered high mass star formation. 
Adams \& Myers~(\cite{AM}) proposed that probably a significant part of 
field stars have formed in small groups consisting of less than a hundred members. 
Studies of star forming regions like L\,1340 
therefore add important pieces of information to the star formation 
history of our Galaxy.
 
In order to find the distribution of high density gas closely 
related to star formation, and its connection to the observed 
signposts of recent star formation we observed the (1,1) and 
(2,2) inversion lines of ammonia in regions of the cloud 
which have shown high column densities in the C$^{18}$O emission. 
The  NH$_3$(1,1) line is suitable for probing the density regimes 
around $n\mathrm{(H_2)} \approx 10^4$\,cm$^{-3}$ (Harju, Walmsley 
\& Wouterloot~\cite{HWW}, hereafter HWW). These densities are high 
enough to shield the gas from interstellar UV radiation, and 
thus disconnect them from the magnetic fields. Ammonia cores are 
those regions of the molecular clouds, where the self-similar structure
resulted from interstellar turbulence breaks down and star formation becomes possible.

In this paper we present the results of the ammonia observations. 
On one hand, we expect to get a better view of the structure of the 
highest density parts of L\,1340, owing to the higher angular resolution of 
our observations than those presented in Paper~I. 
Comparison of physical parameters derived from NH$_3$, C$^{18}$O and 
$^{13}$CO observations, on the other hand, may reveal a relationship 
between different density regimes of the molecular gas.
The ammonia database by Jijina, Myers \& Adams
(\cite{JMA}, hereafter JMA) makes it possible to compare L\,1340
with other star forming regions.
A spectroscopic and photometric follow-up study of the 
candidate YSOs associated with the cloud  will be published 
in a subsequent paper (Kun~\cite{K2002b}). 
 
We describe our ammonia observations  in Sect.~\ref{Sect_2}. The methods 
of our data analysis are presented in Sect.~\ref{Sect_3}, and 
the results of observations in Sect.~\ref{Sect_4}.
In Sect.~\ref{Sect_5} we discuss the connection of ammonia cores 
with candidate YSOs and with their environment revealed by $^{13}$CO, 
C$^{18}$O, \ion{H}{i} and visual extinction. We also compare 
the dense cores of L\,1340  with other star forming clouds.
Sect.~\ref{Sect_6} gives a brief summary of our results. 
Appendices A and B briefly outline how the use of publicly available 
\ion{H}{i} and star count data contribute to get a coherent picture of the cloud.

\section{Observations} 
\label{Sect_2} 
 
We mapped the (1,1) and (2,2) inversion transition lines of  
ammonia towards L\,1340  using the 100-m radio telescope 
of MPIfR at Effelsberg 
in February and October 1997. The half-power beam width of the 
telescope at 23.7~GHz is 40$\arcsec$, 
corresponding to 0.12\,pc at the distance of L\,1340. The facility 1.3\,cm 
maser receiver was used with a typical system temperature of 90\,K. 
The spectrometer was a 1024-channel autocorrelator 
split into two bands of 6.25\,MHz in order to observe simultaneously 
at the frequencies of NH$_3$(1,1) (23694.495\,MHz)  and NH$_3$(2,2) 
(23722.633\,MHz). The spectral resolution was 
0.15\,kms$^{-1}$. 
 
The mapping was done in total power mode with 3 ONs per OFF on a 
40\arcsec\  grid, with additional positions towards the C$^{18}$O 
peaks and optically invisible IRAS point sources. The integration 
times were 3 minutes per position.
We calibrated our measurements using continuum scans of NGC\,7027 
for which we adopted $T_\mathrm{MB}$=8.2\,K, 
corresponding to 5.86\,Jy (Baars et al.~\cite{BGPW}). 
The pointing was checked each 2--3 hours on nearby 
continuum sources.  Its accuracy was about 5$\arcsec$. The typical 
rms noise in our spectra is 0.15\,K ($T_\mathrm{MB}$). 
 
\begin{figure*}
\centering
\includegraphics[width=12cm]{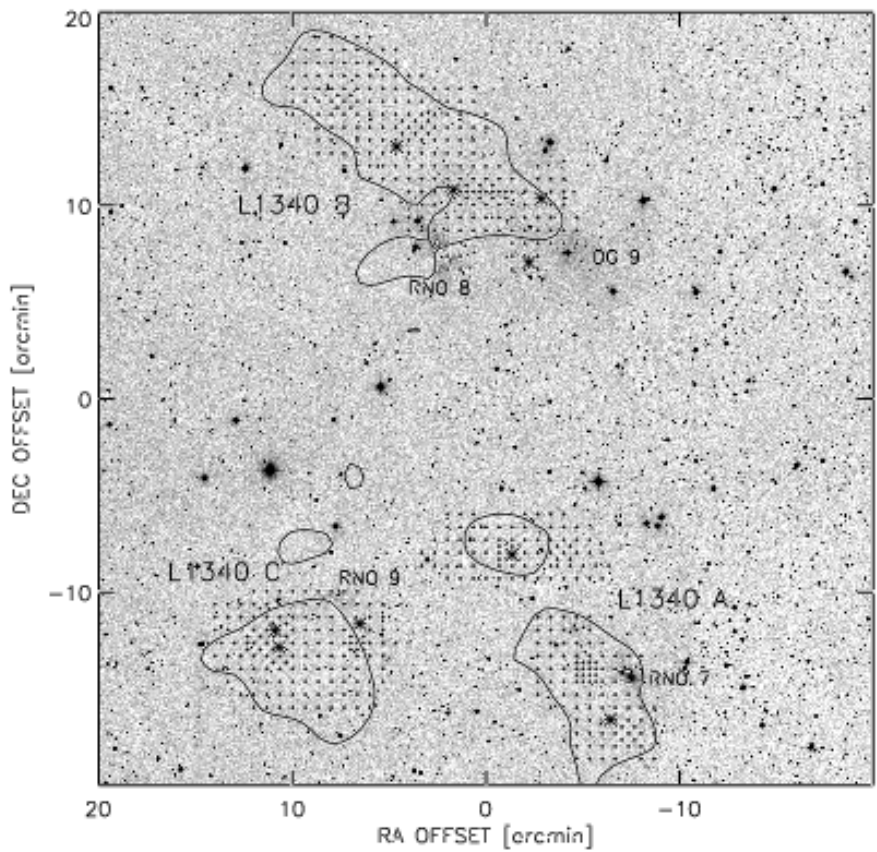} 
\caption{The observed points (crosses) overlaid on the optical image of 
L\,1340 taken from the DSS-1 red.  The C$^{18}$O cores are indicated 
by their $\int\!T_A\,dv = 0.45\,\mathrm{K\,km\,s}^{-1}$ contour. 
Asterisks show the positions of optically invisible and nebulous  
IRAS  point sources. Positions of DG~9 and the RNOs are also 
indicated. Coordinate offsets are given in arcmin with respect to 
RA(2000)=2$^\mathrm{h}30\fm0$, D(2000)=72\degr52\farcm0.} 
\label{Fig1} 
\end{figure*}

Our ammonia observations covered an area of 170~arcmin$^2$, 
or 5.15~pc$^2$ at a distance of 600~pc. 
Virtually the whole area was observed where $\int\!T_A(\mathrm{C^{18}O})\,dv$
(Paper~I) was higher than about 0.45\,K\,km\,s$^{-1}$. 
Figure~\ref{Fig1} shows the observed  positions 
overlaid on the red optical image of L\,1340  
obtained from DSS-1\footnote{Based on photographic data of the National Geographic 
Society -- Palomar Observatory Sky Survey (NGS-POSS) obtained using the Oschin 
Telescope on Palomar Mountain.  The NGS-POSS was funded by a grant from the National
Geographic Society to the California Institute of Technology.  The
plates were processed into the present compressed digital form with
their permission.  The Digitized Sky Survey was produced at the Space
Telescope Science Institute under US Government grant NAG W-2166.}. 
 
\section{Data analysis}
\label{Sect_3}

\subsection{Physical properties of the ammonia gas}
\label{Sect_3.1}
 
Our ammonia spectra were reduced and analysed using the CLASS software
(Forveille et al.~\cite{CLASS}). 
We derived the physical parameters of the NH$_3$ gas following the 
procedure described by HWW. During this procedure  we took into account 
the beam filling factor. It was estimated from the half-maximum sizes 
of the structures in the integrated intensity maps (Figs. 3--5, Sect.~\ref{Sect_4}) 
and the HPBW of the telescope using Martin's \& Barrett's~(\cite{MB}) Eq.~7. 
We obtained average beam filling factor of 0.5, which value we used
in the data analysis. 

The rotational temperature  $T_\mathrm{12}$ can be determined from the (1,1) 
and (2,2) brightness temperatures (Ho \& Townes~\cite{HT}).
At most positions of L\,1340, however, the (2,2) line was too weak 
($T_\mathrm{B}(2,2) < 2\,\sigma$), therefore we averaged the (1,1) and (2,2) 
spectra for the  regions around $T_\mathrm{B}$(1,1) maxima
in order to enhance the S/N and determined $T_\mathrm{12}$ from the
averaged spectra, using Ho \& Townes~(\cite{HT}) Eq.~4. The 
$T_\mathrm{12}$ values were then transformed into kinetic temperatures 
$T_\mathrm{k}$ according to Walmsley \& Ungerechts~(\cite{WU}), using
collisional rate coefficients given by Danby et al.~(\cite{DFVSW}).

The excitation temperature $T_\mathrm{ex}$ of the transition was 
determined where $T_\mathrm{B}(1,1) > 5\,\sigma$, the signal-to-noise 
ratio allowed us to fit the (1,1) spectra with the 
pattern of the hyperfine components. We obtained excitation temperatures 
of 5--6\,K for these positions, therefore we 
assumed $T_\mathrm{ex}=$\,6\,K for the parts of the 
clouds where the S/N of the spectra was insufficient 
for the determination of $T_\mathrm{ex}$. 

The column density $N({\rm NH_3(1,1)})$  was computed using different 
assumptions depending on line intensities in accordance with the criteria 
given by HWW, and total ammonia column densities, 
$N$(NH$_3$), were derived  using HWW's Eq.~7. 

\subsection{Physical properties of ammonia cores} 
\label{Sect_3.2}

The distribution of NH$_3$(1,1) integrated intensity shows several 
peaks, that define the positions of the dense cores of L\,1340.
We defined the ammonia cores as closed areas bordered by the lowest 
significant (3\,$\sigma$\,=0.45\,K\,km\,s$^{-1}$) contours in the NH$_3$(1,1) 
integrated intensity maps. Multiple peaks inside a closed contour are regarded
as different cores if they are separated below the half-maximum contour.
Table~\ref{Tab1} shows the observed properties of the ammonia 
cores. The following quantities are listed: Column~1: the name of the core; 
Cols.~2--3: the offsets of the peak position of the integrated 
intensity in arcmin with respect to  RA(2000)=2$^{\rm h}29^{\rm m}41\fs64$
and Dec(2000)=+72\degr43\arcmin22\farcs2;  Col.~4: the radial velocity obtained 
by fitting  the NH$_3$ hyperfine structure to the observed spectra, 
averaged for the points within the border of the core, and its standard deviation; 
Col.~5: the average line width of the clump and its dispersion; 
This quantity was derived from the NH$_3$(1,1) fit to the mean 
spectra, obtained by averaging individual spectra  within the 3$\sigma$ 
contour of the integrated intensity maps. $\Delta v$ was corrected 
for the spectral resolution. Col.~6: the full angular size of the core within 
the 3\,$\sigma$ contour of the integrated intensity map in arcsec,
not corrected for the beam size. We obtained 
this information by approximating the $3\,\sigma$ contour with an ellipse; 
Col.~7: the angular distance of the nearest  IRAS  source 
from the ammonia peak in arcsec; Col.~8: the name and type of 
the nearest IRAS source. An `s' following the  IRAS  
name marks sources associated with visible stars, and `i' is for optically 
invisible sources.

In this section we describe the methods used for deriving sizes,
kinetic temperatures, hydrogen column densities, masses, and 
nonthermal velocity dispersions of ammonia cores. We estimate their stabilities,
and define their connection to IRAS point sources. 

\begin{table*} 
\centering
\caption{Measured parameters of NH$_{3}$ cores of L\,1340 } 
\label{Tab1} 
\begin{flushleft} 
\begin{tabular}{lcccccrl} 
\noalign{\smallskip} 
\hline\hline
\noalign{\smallskip} 
Core & RA offset & D offset & $v_{LSR}$ & $\Delta v$  
& Ang. size & d(*) & Nearest  \\ 
\noalign{\smallskip}
     &  (\arcmin) & (\arcmin)  &  (km\,s$^{-1}$) & (km\,s$^{-1}$) 
& ($\arcsec \times \arcsec$) & $\arcsec$ &  IRAS  source \\
\noalign{\smallskip} 
\hline
\noalign{\smallskip} 
A1 & $-$6.00 & $-$8.33 & $-$14.13\,(0.03) & 0.64\,(0.06) & 140$\times$60 & 60 & 02238+7222 {\it i} \\ 
A2 & $-$4.00 & $-$5.33 & $-$14.35\,(0.03) & 0.36\,(0.04) & 80$\times$20 &  \\
A3 &  $-$2.33 & ~0.00 & $-$14.56\,(0.01) & 0.28\,(0.02) & 70$\times$60 \\ 
A4 &  ~0.33 & ~0.33 & $-$13.75\,(0.30)  & 0.99\,(0.08) & 90$\times$60 & 10 &  02249+7230 {\it i} \\ 
B1 &  ~2.33 & 19.67 &  $-$14.37\,(0.15) &  0.63\,(0.03) & 85$\times$70 & 40 & F02256+7249 {\it i} \\ 
B2 &  ~4.00 & 20.00 & $-$14.67\,(0.03) & 0.29\,(0.05) & 90$\times$40 & 68  \\
C1 & ~8.33 & $-$4.33 & $-$14.63\,(0.09) & 0.67\,(0.04) & 100$\times$70 & 80 & 02267+7226 {\it i} \\  
C2 & 12.33 & $-$3.00 & $-$15.89\,(0.14) & 0.84\,(0.04) & 80$\times$70 & 20 & F02277+7226 {\it i} \\ 
C3w & 12.67 & $-$4.67 & $-$15.71\,(0.36) & 0.97\,(0.05) & 100$\times$60 & 50 & 02276+7225 {\it i} \\ 
C3e & 13.67 & $-$4.00 & $-$15.92\,(0.10) & 0.40\,(0.03) & 80$\times$50 & 40 & F02279+7225 {\it s} \\[4pt] 
\noalign{\smallskip}
\hline 
\end{tabular}
\end{flushleft}
\end{table*}

The {\em half-maximum radii\/} $R_{1/2}$ of the cores, corrected for the angular 
resolution of the observations were derived as  
$R_{1/2} =[A _{1/2}/\pi- (\mathrm{HPBW}/2)^{2}]^{1/2}$, where $A_{1/2}$ was the area of 
the core within the half-maximum contour of the integrated intensity maps. 

\begin{figure}
\resizebox{\hsize}{!}{\includegraphics{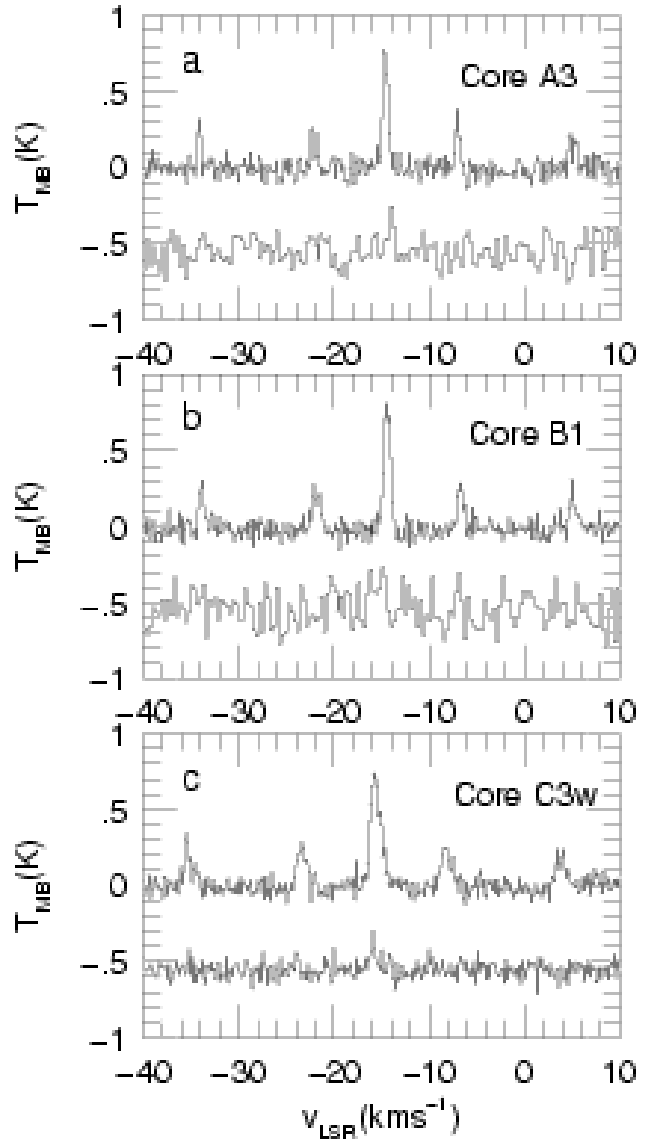}}
\caption{Averaged spectra used for determining $T_{12}$ for
cores a) A3, b) B1, c) C3w.}
\label{Fig2}
\end{figure}

In order to determine the {\em mean kinetic temperatures\/} of the cores we
averaged the (1,1) and (2,2) spectra over the positions within the lowest
significant integrated intensity contours. 
The resulting spectra, due to their higher S/N made it 
possible to measure the integrated intensity of the  (2,2) lines.
Figure~\ref{Fig2} shows the average spectra for cores {\it A3\/}, {\it B1\/}
and {\it C3w\/}. 

Having estimates on the kinetic temperatures the {\em nonthermal
line widths\/} of the cores can be determined:
$\Delta v_\mathrm{NT}^2 = \Delta v_\mathrm{obs}^2 - 8\ln 2kT_\mathrm{k} / m_\mathrm{obs},$
where $\Delta v_\mathrm{NT}$ is the FWHM of the nonthermal (turbulent) motion, and
$m_\mathrm{obs} = 17\,m_{H}$ is the mass of the ammonia molecule.

{\em Hydrogen column densities\/} $N$(H$_2$) were calculated from ammonia column 
densities with the assumption that 
NH$_3$/H$_2 = n\mathrm{(NH_3)}/n(\mathrm{H}_2)=3\times10^{-8}$,
a mean value predicted by the models of chemical evolution of protostellar
cores (Bergin \& Langer~\cite{BL}). According to the models NH$_3$/H$_2$ is 
constant during the protostellar collapse, but increases during 
the later evolution of cores, when shocks accompanying star formation
release the molecules formed on grains (Nejad et al.~\cite{NWC}; 
Suzuki et al.~\cite{SYOKIHT}). 
For those positions of the cloud, where both $T_\mathrm{ex}$ 
and $T_\mathrm{k}$ were determined, $n$(H$_2$) was derived 
independently of NH$_3$/H$_2$  (Ho \& Townes~\cite{HT}, Eq.~2). 
These positions allow us to check the adopted value of NH$_3$/H$_2$ 
when we make an assumption on the geometry of the cores.

{\em Core masses\/}  were determined by summing up the mass elements 
$N(\mathrm{H}_2)dx dy$ for each position with 
$N\mathrm{(NH_3)} \geq 1.2\times\,10^{14}$\,cm$^{-2}$,
the lowest closed contour for most of the cores, and adding 20\%
helium (in number density). Here $N(\rm {H}_2)$ is the 
hydrogen column density and $dx dy$ is the area 
corresponding to the grid spacing at a distance of 600\,pc. 
We observed at a 40\arcsec\ grid, with several additional 
points halfway between two grid points, therefore we interpolated linearly
the column densities so that we got a regular grid with spacing of 20\arcsec. 

In estimating the dynamical states of the ammonia cores their gravitational 
and kinetic energies, as well as the external pressure due to the 
weight of the overlying cloud have to be taken into account. 
Inserting the  mass and radius of L\,1340  
obtained from $^{13}$CO data (Yonekura et al.~\cite{YDMOF}) into the formula 
$P_\mathrm{ext}\,/\,k \approx 1.4 (M/M_{\sun})^{2}(R/\mathrm{pc})^{-4}$
(Bertoldi \& McKee~\cite{BMK}) resulted in the mean external pressure 
$P_\mathrm{ext}\,/\,k \approx 2.4\times 10^{5}$\,K\,cm$^{-3}$. With this data 
we estimated the critical Bonnor--Ebert mass of the cores,
$M_\mathrm{BE}=1.18\,\sigma^{4}/(G^{3}\,P_\mathrm{ext})^{1/2}$,
\noindent where  $\sigma^{2}=\Delta v^{2}/8\,\ln 2$, and
$G$ is the gravitational constant. Cores more massive than $M_\mathrm{BE}$
will collapse to form stars in this environment, if other effects
are neglected.

\subsection{Connection of cores with YSOs}
\label{Sect_3.3}

Optically invisible IRAS point sources projected on the cores are 
supposed to be embedded YSOs born in the cores. We associate an embedded YSO with a
core, following JMA, if it lies within 2$\times R_{1/2}$ to the peak of the core.
Connection of optically visible YSOs with the cores is less obvious.
These more evolved objects may either have been born in the core on which
they are projected or in another core which has already dispersed.
In the following discussions we shall assume the cores having 
optically identified IRAS point sources or H$\alpha$ emission
stars within 2$\times R_{1/2}$ to their peak $T_\mathrm{B}(1,1)$
to be starless. Our classification is somewhat uncertain because, 
due to the large distance of L\,1340, only the most luminous 
members of the YSO population were detected by IRAS. Most IRAS 
sources in L\,1340 are faint, close to the detection threshold, 
have less than four good quality fluxes, and  are extended in 
the 100\,$\mu$m image. We note that only one optically invisible source, 
IRAS 02249+7230 has a good flux quality at 100\,$\mu$m.

\section{Results}
\label{Sect_4}

\subsection{Distribution of ammonia in L\,1340}
\label{Sect_4.1}

\begin{figure*} 
\centering{
\includegraphics[width=8.0cm]{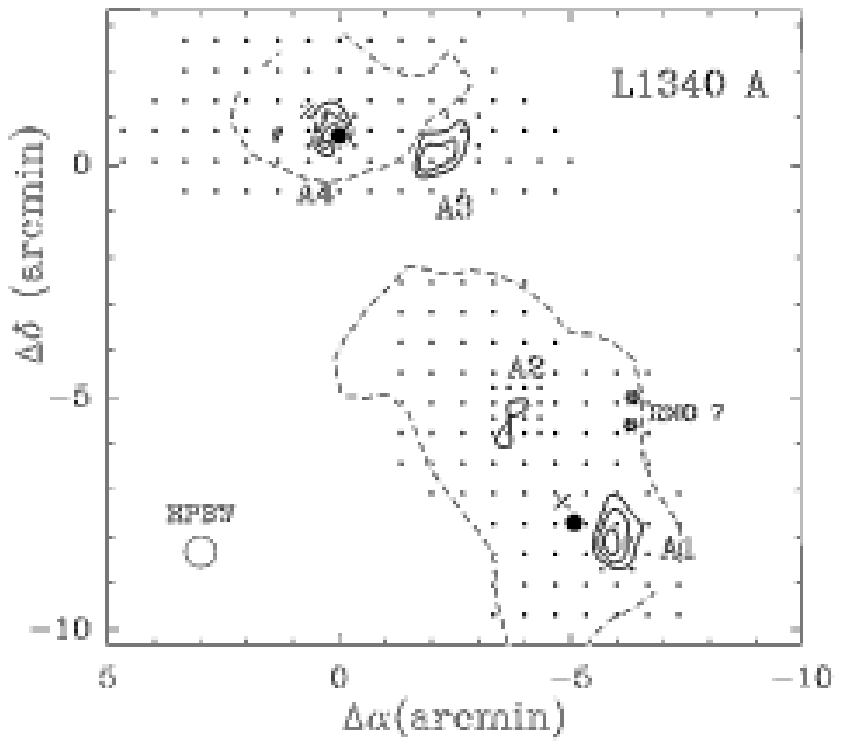}
\includegraphics[width=8.0cm]{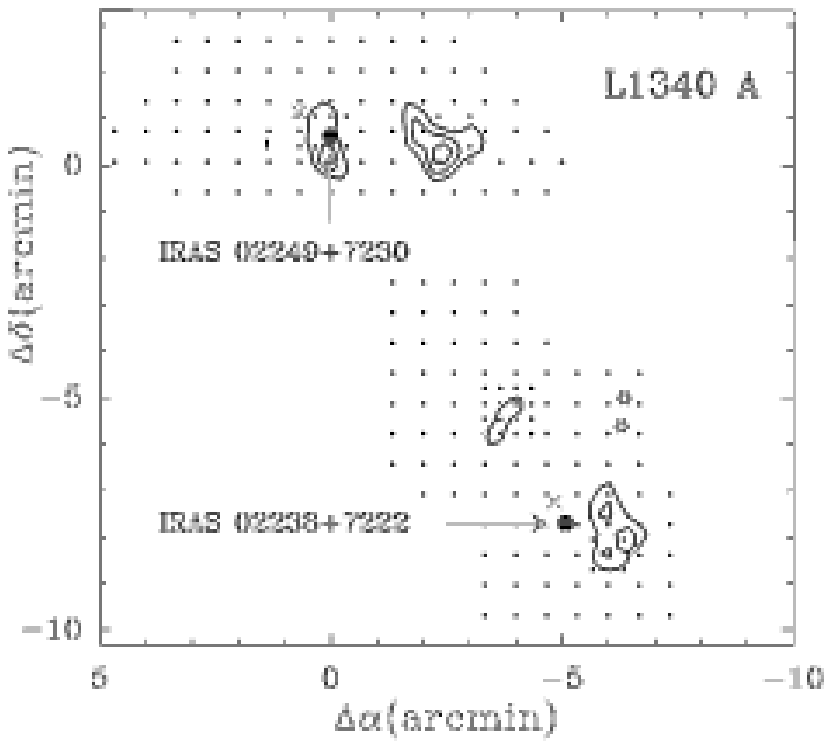}}
\caption{{\em left\/}: NH$_3$(1,1) (solid contours) main group integrated intensity map 
of L\,1340~A. The emission was integrated over the velocity 
interval between $-$17\,km\,s$^{-1}$ and $-$12\,km\,s$^{-1}$.
The lowest contour is 0.40\,K\,km\,s$^{-1}$, and the 
increment is 0.15\,K\,km\,s$^{-1}$. The observed positions are shown by dots.
Black circles mark the positions of optically invisible  
IRAS point sources, and asterisks indicate those associated with visible 
stars. Crosses show the positions of the C$^{18}$O peaks. 
Dashed contours indicate the 0.45\,K\,km\,s$^{-1}$ level 
of the $\int\!\mathrm{C^{18}O}\,dv$ distribution. The HPBW of the ammonia observations
is shown in the lower left corner. The HPBW of the C$^{18}$O was 2\farcm7.
Coordinate offsets are given in arcmin with respect to 
RA(2000)=2$^{\rm h}29^{\rm m}41\fs64$ and Dec(2000)=+72\degr43\arcmin22\farcs2.
{\em right\/}: Ammonia column density map of 
L\,1340~A. The lowest contour is at 1.2$\times 10^{14}$ cm$^{-2}$, and the 
increment is 0.6$\times 10^{14}$ cm$^{-2}$.}
\label{Fig3} 
\end{figure*}

\begin{figure*} 
\centering{
\includegraphics[width=8.0cm]{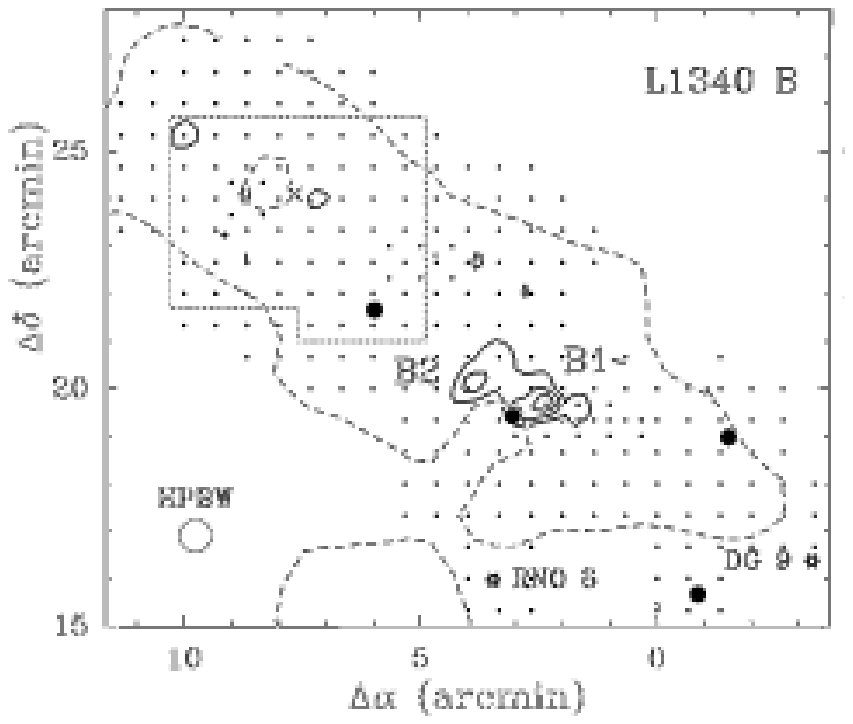}
\includegraphics[width=8.0cm]{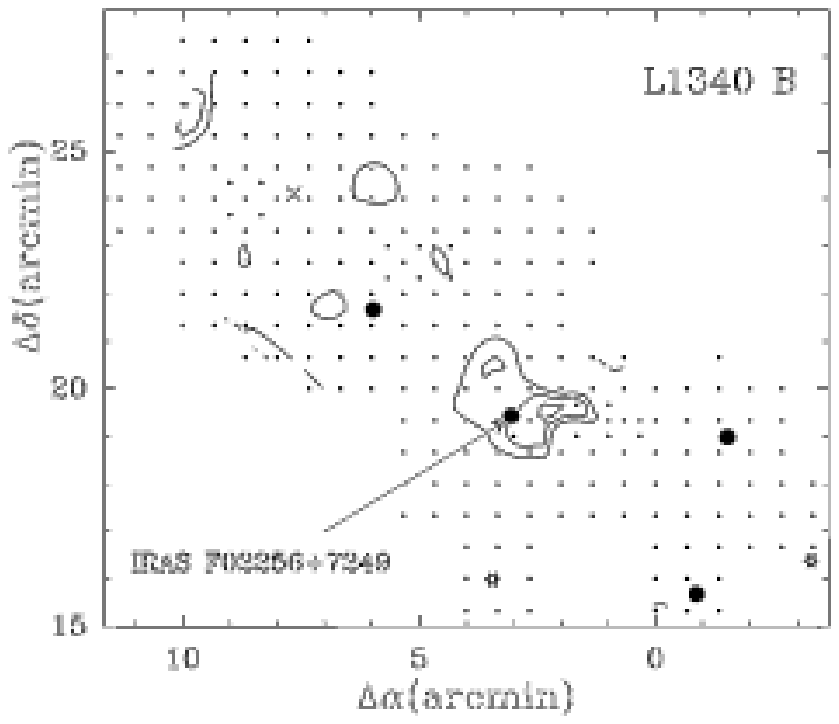}}
\caption{Same as Fig. 3, but for L\,1340~B. {\em left\/}: Contours start 
at 0.40\,K\,km\,s$^{-1}$, and the increment is 0.20\,K\,km\,s$^{-1}$. The polygon
drawn by dotted line indicates an extended region where weak emission 
($T_\mathrm{B}(NH_3(1,1)) \leq 2\,\sigma$) was detected.
{\em right\/}: The lowest contour is at 1.2$\times 10^{14}$ cm$^{-2}$, and the 
increment is 0.4$\times 10^{14}$ cm$^{-2}$.} 
\label{Fig4} 
\end{figure*}

\begin{figure*} 
\centering{
\includegraphics[width=8.0cm]{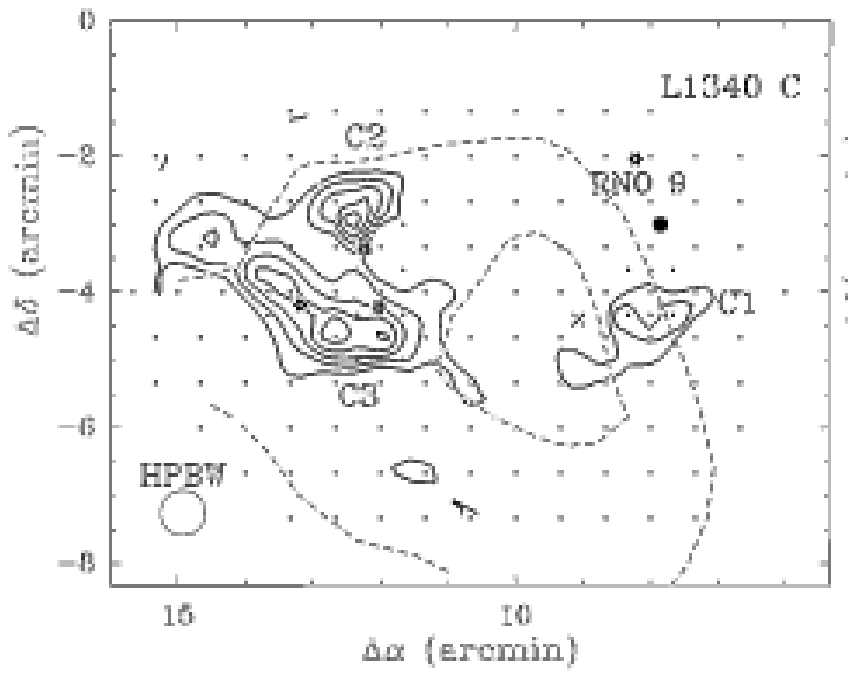}
\includegraphics[width=8.0cm]{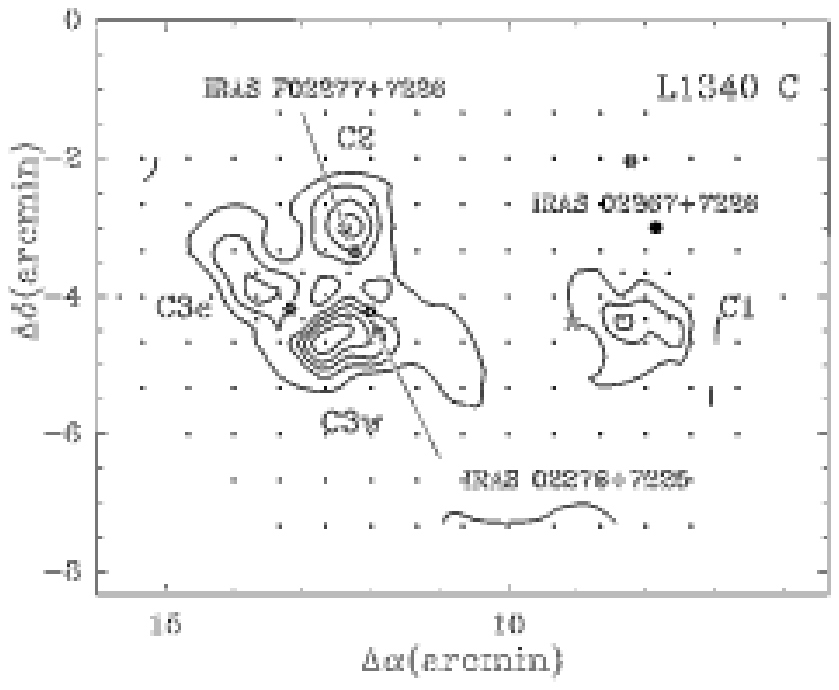}}
\caption{Same as Fig. 3, for L\,1340~C. {\em left\/}: Contours start 
at 0.45\,K\,km\,s$^{-1}$, and the increment is 0.20\,K\,km\,s$^{-1}$. 
Due to the larger velocity range observed in this clump the emission was 
integrated  over the velocity interval of $-$19\,km\,s$^{-1}$--$-$12\,km\,s$^{-1}$.
{\em right\/}: The lowest contour is at 1.2$\times 10^{14}$ cm$^{-2}$, and the 
increment is 0.8$\times 10^{14}$ cm$^{-2}$.}
\label{Fig5} 
\end{figure*}

The ammonia cores defined by the integrated intensity distribution of the 
main-group of the (1,1) line are shown in the left panels of 
Figures~\ref{Fig3}--\ref{Fig5} for clumps {\it A\/}, {\it B\/}, and 
{\it C\/}, respectively. For comparison, we also plotted 
the C$^{18}$O contours $\int\! T_A\,dv$= 0.45\,K\,km\,s$^{-1}$ and
0.75\,K\,km\,s$^{-1}$. The ammonia cores are
labelled in the figures. 

\begin{figure}
\resizebox{\hsize}{!}{\includegraphics{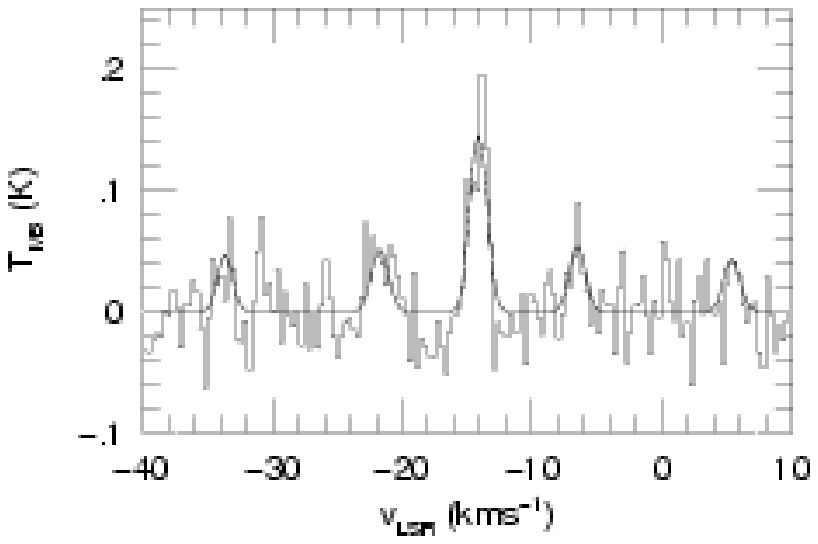}}
\caption{Average of 56 NH$_3$(1,1) spectra in the region centred on the C$^{18}$O peak
position of clump {\it B\/}.}
\label{Fig6}
\end{figure}

In addition to the cores defined in Sect.~\ref{Sect_3.2} there is an 
extended region of weak NH$_3$(1,1) 
emission in the northern part of clump {\it B\/}, around the position of the 
C$^{18}$O peak. The integrated intensity of the (1,1) line
is below the 3$\sigma$ limit at most positions. The optically invisible 
source IRAS 02263+7251 lies in this area. Averaging 
56 spectra around the position the C$^{18}$O peak (bordered by a dotted polygon
in Fig.~\ref{Fig4}) we obtained the spectrum displayed in Fig.~\ref{Fig6}. The 
weak line indicates low average column density for this region. Because the critical 
density of the excitation of NH$_3$(1,1) emission is about
10$^4$\,cm$^{-3}$, this part of the cloud probably contains high density 
regions much smaller than the angular resolution of our observations.

Column density maps are shown in the right panels of Figs.~\ref{Fig3}--\ref{Fig5}.
IRAS point sources associated with the cores are labelled in these figures.
Because of the effect of the optical depth, column densities 
are not directly proportional to the integrated intensities.
Comparison of the two sets of maps shows the main structures to be 
largely similar, with the exception that core~{\it C3\/} splits into two parts, 
{\it C3w\/} and {\it C3e\/}, in the column density map.

The physical properties of the cores, derived by the procedures described  
in Sect.~\ref{Sect_3.2}, are displayed in Table~\ref{Tab2}. The  following 
quantities are listed: 
Col.~1: name of the core. An asterisk following the name indicates that we 
associated the core with an embedded YSO; Col.~2: the half-maximum radius $R_{1/2}$, 
in parsecs; Col.~3: $T_\mathrm{ex}$  at the peak position 
where the S/N of the line allowed its determination; Col.~4: 
the mean kinetic temperature $T_\mathrm{k}$; Col.~5: the nonthermal
component of the line width $\Delta v_\mathrm{NT}$;
Col.~6: the maximum column density $N_\mathrm{max}$(NH$_3$); Col.~7:
volume density $n$(H$_2$) of the hydrogen derived from $T_\mathrm{ex}$;
Col.~8: the mass of the core in solar masses. The Bonnor--Ebert mass is shown in Col.~9.
Bolometric luminosity of the optically invisible  IRAS  point 
source associated with the core, calculated from the IRAS 
fluxes adding the long-wavelength bolometric correction 
(Myers et al.~\cite{MFMBBSE}) is shown in Col.~10. Where only flux upper limits were 
available, we estimated the fluxes from the infrared data sets (IRDS) obtained
via the {\it IRAS Software Telescope\/} maintained at SRON (Assendorp et al.~\cite{ABJKRW}).

\begin{table*}
\centering
\caption{Derived physical parameters of the NH$_{3}$ cores of L\,1340 } 
\label{Tab2}
\begin{flushleft}
\begin{tabular}{lccc@{\hspace{2mm}}c@{\hspace{2mm}}
c@{\hspace{2mm}}cc@{\hspace{2mm}}c@{\hspace{3mm}}r} 
\hline\hline
\noalign{\smallskip}
Core & $R_{1/2}$ & $T_\mathrm{ex}$ & $T_\mathrm{k}$ & $\Delta v_\mathrm{NT}$ &
$N_\mathrm{max}$(NH$_3$) & $n\mathrm{(H_2)}$ & $M\mathrm{(NH_3)}$ &  $M_\mathrm{BE}$ & $L_\mathrm{IRAS}$ \\ 
\noalign{\smallskip} 
\cline{3-4} \cline{8-9}   \\[-8pt]
\noalign{\smallskip}
     & (pc) & \multicolumn{2}{c}{(K)} & (km\,s$^{-1}$) & (10$^{14}$cm$^{-2}$) 
      & (10$^{4}$cm$^{-3}$)  & \multicolumn{2}{c}{(M$_{\sun}$)}  & (L$_{\sun}$) \\
\noalign{\smallskip}
 \hline\noalign{\smallskip}
A1* & 0.08 &  $\cdots$ & ~15.2\,(2.0) & 0.61 & 2.34\,(0.24) & $\cdots$ & 5.8 & ~3.2  &  $\leq$9.6 \\ 
A2 & 0.04 &  $\cdots$ & $\leq$12.5 & 0.31 & 2.10\,(0.50) & $\cdots$ & 1.5 & ~0.4  \\
A3 & 0.08 &  4.7  & ~11.9\,(2.3) & 0.21 & 3.40\,(0.33) & 0.99\,(0.08) & 6.5 & ~0.2   \\  
A4* & 0.06 &  $\cdots$ &  ~13.5\,(2.0) & 0.97 & 2.53\,(0.60) & $\cdots$ & 3.7 & 18.6 &  4.9 \\ 
B1*  & 0.10 &  5.2 & ~14.6\,(3.1) & 0.67 & 2.05\,(0.27) & 1.44\,(0.10) & 5.6 & ~3.0 & 8.8 \\ 
B2  & 0.08 & $\cdots$ & $\leq$15.0 & 0.24 & 1.45\,(0.40) & $\cdots$ & 2.7 & ~0.2  \\
C1*  & 0.10 & 4.7  & ~14.1\,(1.7) & 0.64 & 3.57\,(0.46) & $\cdots$ & 12.0 & 3.9  & 2.8 \\ 
C2*  & 0.10 & 4.6  & ~16.7\,(1.6) & 0.81 & 4.61\,(0.60) & 1.07\,(0.18) & 10.2 & ~9.7 & $\cdots$  \\ 
C3w* & 0.15 & 5.2  & ~13.6\,(1.1) & 0.95 & 4.89\,(1.50) & 1.82\,(0.40) & 15.6 & 17.2 & 1.5 \\ 
C3e & 0.11 & 6.0  & ~12.5\,(4.0) & 0.36 & 3.73\,(0.57) & 1.29\,(0.36) & 15.4 & ~0.5  \\
\noalign{\smallskip}
\hline
Mean & 0.08 & 5.1 & 14.0 & 0.58 & 3.07 & 1.32 & 7.9 & 5.7  & 5.4 \\
Starless cores &  0.07 & 5.1 & 12.2 & 0.28 & 2.67 & 1.14 & 6.5 & 0.3 & $\cdots$ \\
Cores with stars & 0.09 & 5.0 & 14.6 & 0.78 & 3.33 & 1.44 & 8.8 & 9.3 & 5.4 \\
\noalign{\smallskip}
\hline 
\end{tabular}
\end{flushleft}
\end{table*} 

The observed ammonia cores probably represent the densest regions
of L\,1340. {\it B1\/} and {\it B2\/}, as well as
{\it C3w\/} and {\it C3e\/} constitute twin core systems according to the 
definition by JMA. The cores are located close to the C$^{18}$O peaks 
in clump {\it A\/} within the accuracy set by the different
angular resolutions. In clumps {\it B\/} and {\it C\/}, however, the
high density regions indicated by the ammonia emission are located
far from the column density peaks of the C$^{18}$O. These small dense 
regions might have been missed during the C$^{18}$O survey because of
their half-maximum sizes are smaller than the grid spacing (2\arcmin). 
The total mass in the dense cores is 79\,M$_{\sun}$, some 6\,\% of the
mass traced by C$^{18}$O. 

\subsection{Velocity structure}
\label{Sect_4.2}

While neither C$^{18}$O nor NH$_3$ observations have indicated velocity gradients 
in L\,1340~A and L\,1340~B, C$^{18}$O measurements of L\,1340~C 
have shown a clear radial velocity gradient of 0.71\,km\,s$^{-1}$\,pc$^{-1}$ 
in the galactic longitude direction, which was interpreted  
as rotation of the clump in Paper~I. The ammonia data, having 
higher angular resolution, suggest another possible scenario.
Figure~\ref{Fig3} shows that clump~C contains two high density
regions, separated by a lower density region between the right ascension offsets of 
about 9\farcm16  and 10\farcm83. The two subclumps have a velocity 
difference of about 1.2\,km\,s$^{-1}$ (Table~\ref{Tab2}). The observed velocity 
gradient may result from the overlapping of the two clumps of different 
radial velocities. A similar situation was found in Orion~KL by 
Wang et al.~(\cite{WWW}).

\begin{figure}
\resizebox{\hsize}{!}{\includegraphics{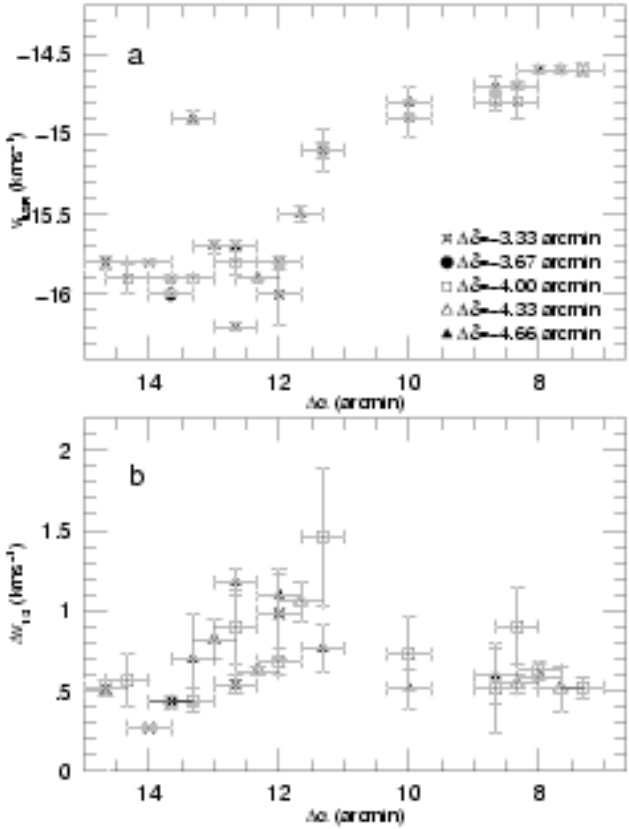}}
\caption{{\bf a)} Radial velocities in Clump~C as a function of RA offset; 
{\bf b)} Linewidths observed in Clump~C as a function of RA offset.
Different declination offsets are marked 
with different symbols. Offsets are given in arcmin with respect to
RA(2000)=2$^{\rm h}29^{\rm m}41\fs64$
and Dec(2000)=+72\degr43\arcmin22\farcs2.}
\label{Fig7}
\end{figure}

Figure~\ref{Fig7}a displays $v_\mathrm{LSR}$  as a function
of $\Delta\,\alpha$, at several $\delta$ offsets. The less negative velocity
component at $\Delta \alpha < 11\arcmin$ shows a small velocity gradient. 
The velocity changes abruptly between the offsets 11\farcm33 and $12\farcm00$,
and is nearly constant (about $-$15.6\,km\,s$^{-1}$)  at larger offsets.  
Both components can be observed  at $11\farcm33 \leq \Delta \alpha \leq 13\farcm33$.
This overlapping shows up as an increase in the line widths in this 
$\Delta\,\alpha$ interval (Fig.~\ref{Fig7}b). The region of enhanced 
line widths coincides with the part of the clump where IRAS 
point sources are found. This morphology suggests that clump collision
might have played role in triggering star formation in L\,1340~C.

\section{Discussion}
\label{Sect_5}

Ammonia cores are those regions of the interstellar medium where the volume 
density is between 10$^4$--10$^5$\,cm$^{-3}$. 
In respect of physical properties they do not form a homogeneous 
group. Starless cores in most clouds have smaller nonthermal 
line widths and masses, and lower kinetic temperatures than those 
associated with IRAS sources. Moreover, the cores associated with embedded or 
nearby young clusters are the most massive and most turbulent 
objects in JMA's ammonia data base. The typical mass and turbulent energy 
of gas in the cores, moreover, varies from cloud to cloud.
Most cores in Taurus form isolated stars, whereas some of them in Ophiuchus
and Orion give birth to rich clusters (Motte et al.~\cite{MAN}; 
Mitchell et al.~\cite{MJMFT}). Several observational and theoretical
studies suggest that the observed nonthermal line widths of
cores are related to the initial conditions of star formation
(e.g. Myers \& Fuller~\cite{MF}; Caselli \& Myers~\cite{CM}).

Dense cores of molecular clouds are thought to be created by shocks due to
the supersonic turbulent velocity field of the ISM, referred to as
turbulent fragmentation (e.g. Elmegreen~\cite{BGE}; Klessen, Heitsch \& 
Mac Low~\cite{KHM}).
In this scenario several observed properties of core/cloud systems are 
related to the nature of interstellar turbulence
(Padoan~\cite{Padoan}; Padoan \& Nordlund~\cite{PN}). In particular, the slope 
$\alpha$ of the line width--size relation 
$\log \Delta v_\mathrm{NT} \propto \alpha\,\log R$ reflects the power spectrum  
of the turbulence, so that $E(k) \propto k^{-\beta}$ and $\alpha = (\beta-1)/2$. 
Density  ratio between cores and their environment, typical core diameter 
and mass, as well as the volume filling factor of the cores are related to 
the size $L_0$ and velocity dispersion $\sigma_{v,0}$ of the ambient cloud.

Cores of various size and velocity dispersion probably define the smallest 
scale of the self-similar structure of interstellar medium.
In low-mass star forming regions they represent the size scale where the 
nonthermal velocity dispersion becomes subsonic (Goodman et al.~\cite{GBWH}).
Myers~(\cite{Myers}) has shown that the strongly turbulent, massive cores having 
$\Delta v > 0.9$\,km\,s$^{-1}$ and N(H$_2) > 1\times10^{22}$\,cm$^{-2}$ 
may contain several critically stable condensations (kernels) cut off from MHD waves 
due to the high extinction of the core. This model suggests that 
massive, cluster-forming cores also represent an inner scale of the 
self-similar structure.

In this section we attempt to deduce some attributes of star formation from 
the derived properties of ammonia cores of L\,1340 (Sect.~\ref{Sect_5.1}), 
compare the features revealed by different tracers with each other  
(Sect.~\ref{Sect_5.2}), and  L\,1340 with other star forming 
regions (Sect.~\ref{Sect_5.3}).

\subsection{Connection of ammonia cores with star formation}
\label{Sect_5.1}

Table~\ref{Tab2} shows that $M$(NH$_3$) $\ge M_\mathrm{BE}$ for most of the
cores of L\,1340. Several observations have shown that this is a necessary 
condition of star formation  (Williams et al.~\cite{WBMK}). Thus the observed 
cores probably highlight the positions of present and future star formation.
The cores associated with embedded YSOs clearly differ from the starless
cores in their nonthermal line widths. This is also true for the twin systems.
The mean $\Delta v_\mathrm{NT}$ of cores without embedded or nearby IRAS 
point source, 0.28\,km\,s$^{-1}$, corresponds to a  velocity dispersion 
$\sigma_\mathrm{NT}=0.12$\,kms$^{-1}$. This is smaller than the isothermal 
sound speed at 13\,K, $c_s$=0.21\,kms$^{-1}$. Thus the detected 
starless cores are among the smallest clumps formed by turbulent 
fragmentation. Such objects may have a wide range of mass (e.g.~Padoan \& 
Nordlund~\cite{PN}), including small clumps which do not collapse.
The weak ammonia emission observed at the northern part of clump~B 
probably originates from such small, dense regions.
$M$(NH$_3$) $\gg M_\mathrm{BE}$ for the starless cores, indicating that 
they are destined to collapse. Our observations thus suggest that 
these cores are prestellar. We note, however, that this 
conclusion has some uncertainties. First, magnetic fields, 
neglected here due to lack of data, may modify the critical 
mass so that it will be significantly larger than $M_\mathrm{BE}$.
Furthermore, recent results by Tafalla et al.~(\cite{TMCWC}) demonstrate 
that ammonia abundance is enhanced towards the centres of some starless 
cores. Detection of the central regions only, enriched in ammonia, 
may lead to overestimation of the mass. Finally, it is possible that these 
cores are not starless, but contain low-luminosity embedded YSOs below the 
detection threshold of IRAS. Observations in other molecular lines 
with high angular resolution and more sensitive infrared observations 
can clarify the nature of these cores. 

Cores associated with embedded IRAS sources have average 
$\Delta v_\mathrm{NT} = 0.78$\,km\,s$^{-1}$, comparable to the those of 
Orion~B (JMA). From the six cores, column density  and nonthermal line 
width of {\it A4\/} and {\it C3w\/} fulfil the criteria set by 
Myers~(\cite{Myers}) for cluster forming cores. In core {\it A4\/} 
IRAS 02249+7230 closely coincides with 
the peak intensity of C$^{18}$O, NH$_3$ and $A_\mathrm{V}$. 
$M_\mathrm{BE} \gg M$(NH$_3$) for this core, suggesting that it is 
disrupting. Morphology of HH~489, associated with the IRAS source, however,
indicates that the direction of the bipolar outflow from this star lies close 
to the plane of the sky (Kumar et al.~\cite{KAY}). The large nonthermal 
velocity dispersion of this core thus cannot arise from the interaction of outflow 
with the core gas. It indicates either the presence of other YSOs with 
outflows along the line of sight, or might have been produced before the 
star formation. 
The other cluster-forming core candidate, {\it C3w\/}, has a common
envelope with {\it C2\/} and {\it C3e\/}. Our ammonia 
observations show this core to be the densest region of L\,1340, 
though it lies far from the C$^{18}$O peak, and is associated with 
a single low-luminosity IRAS source IRAS 02276+7225. No outflow, maser 
source or HH-object have been detected around this source.  
Core {\it C3w\/} is probably  less evolved than {\it A4\/}. 

Our observational results  suggest that the large turbulent 
velocity dispersions of IRAS-associated cores cannot be attributed to 
YSO winds. These cores are not simply more evolved versions of the 
starless cores, but probably form more massive stars than their narrow-line 
counterparts, and some of them  will evolve into small stellar groups similar to 
the two sparse young clusters RNO\,7 and RNO\,8, found in L\,1340
(Kumar et al.~\cite{KAY}; Kun~\cite{K2002a}). In order to reveal the real 
nature and evolutionary state of the cores, their detailed density and 
velocity structures and stellar contents have to be studied via higher 
resolution molecular and submillimeter continuum observations.

\subsection{Comparison with \ion{H}{i}, $^{13}\mathrm{CO}$, 
$\mathrm{C^{18}O}$, and  $A_\mathrm{V}$}
\label{Sect_5.2}

The nonthermal line width--size relation for the structures shown by 
different tracers, called {\em Type 3 line width--size relation\/} 
by Goodman et al.~(\cite{GBWH}), is a useful indicator of the overall 
density structure of a cloud, which, in turn, is closely related to 
the mode of star formation. In order to derive this relation for L\,1340 
we supplemented our ammonia results with C$^{18}$O, $^{13}$CO and 
\ion{H}{i} data.

The $\Delta v_\mathrm{NT}$ and $R$ data for the C$^{18}$O and $^{13}$CO
structures were taken from Paper~I and from Yonekura et al.~(\cite{YDMOF}),
respectively. The size and line width of the \ion{H}{i} structure associated 
with L\,1340 were estimated from the Leiden--Dwingeloo \ion{H}{i} survey 
data (Hartmann \& Burton~\cite{HB}). The main properties of the 
neutral hydrogen in the galactic environment of L\,1340 are shown in 
Appendix~A. The \ion{H}{i} spectra in this region show definite peaks in the 
velocity interval $-18\,\mathrm{km\,s}^{-1} < v_\mathrm{LSR} < -8\,\mathrm{km\,s}^{-1}$
whose characteristic FWHM is $7\,\mathrm{km\,s}^{-1}$, and  
the half-maximum size of the interstellar feature delineated by this gas
component is 38\,pc. 

\begin{figure}
\resizebox{\hsize}{!}{\includegraphics{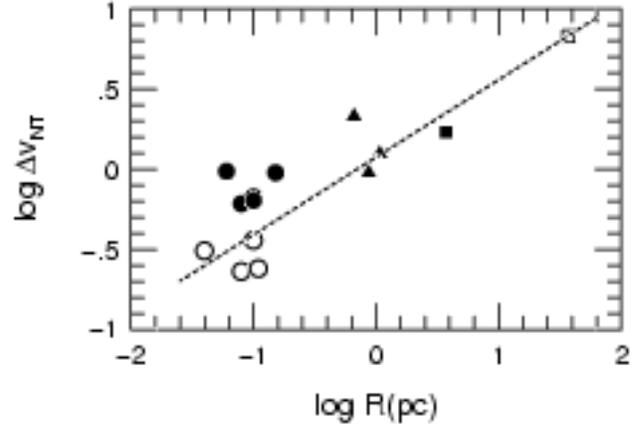}}
\caption{The nonthermal line width--size relation for different substructures 
(ammonia cores, C$^{18}$O clumps and the whole $^{13}$CO cloud) of
L\,1340. Open circles mark the starless NH$_3$ cores and those associated with
optically visible stars, black circles represent the cores associated with 
IRAS point sources. Triangles are for the C$^{18}$O clumps, and black 
square marks the whole $^{13}$CO cloud. The open square shows the \ion{H}{i}
feature, whose half-maximum size was estimated from Fig.~A.2, and in estimating
the nonthermal line width a kinetic temperature 80\,K was assumed.
The dashed line is fitted to all points.}
\label{Fig8}
\end{figure}

The $\log \Delta v_\mathrm{NT}$ vs. $\log R$ relation for the structures 
observed in NH$_3$, C$^{18}$O, $^{13}$CO and \ion{H}{i} is shown in Fig.~\ref{Fig8}.
The $R_{1/2}$ values plotted have been corrected for the different beam sizes
of the observations, and $\Delta v_\mathrm{NT}$ values have been corrected for 
spectral resolutions. We obtained the relation

\begin{equation}
\log \Delta v_\mathrm{NT} = (0.41\pm0.06)\,\log R + (0.12\pm0.06) \label{eq_1}
\end{equation}
\noindent
and the correlation coefficient 0.85.

This relationship reveals the self-similar hierarchy of substructures
from the large \ion{H}{i} cloud to the ammonia cores, i.e. on the 
0.1--40\,pc size scale, suggesting that they are parts of a physically  
connected structure shaped by interstellar turbulence (Larson~\cite{Larson}). The 
slope $\alpha=0.41$ is between those obtained for Taurus 
($0.53\pm0.07$) and Orion~B ($0.21\pm0.03$) cores (Caselli \& Myers~\cite{CM}),
from the same tracers.

We compare properties of NH$_3$ cores and their embedding C$^{18}$O 
clumps in Table~\ref{Tab5}. The data listed show that
the average density ratio of the cores and their embedding clumps $n_c/n_0$,
the typical core diameter $l_c$ and the volume filling factor of the cores
are in accordance with the values predicted by the model of turbulent
fragmentation (Padoan~\cite{Padoan}; Padoan \& Nordlund~\cite{PN}). 
The size and velocity dispersion of the $^{13}$CO cloud 
are $L_0$=3.7\,pc and  $\sigma_{v,0}=0.72$\,km\,s$^{-1}$, 
respectively, thus the large-scale Mach number is 
$\mathcal{M}_{0}=\sigma_{v,0}/c_s =3.4$. With these values the model 
gives $n_c/n_0 \approx \mathcal{M}_{0}^2=11.6$, in accordance with 
the observed $n_c/n_0 \approx 10$. The typical core diameter,
$l_c \sim L_0\,\mathcal{M}_{0}^{-1/\alpha}=0.25$\,pc, is also 
comparable to the observed average 0.16\,pc. The volume filling 
factor of the cores, obtained from the probability density 
function of $n_c/n_0$, is 0.02, compatible with the observed 
average shown in Table~\ref{Tab5}.

\begin{table}
\centering 
\caption{Comparison of C$^{18}$O and NH$_3$ cores in L\,1340 } 
\label{Tab5} 
\begin{flushleft} 
\begin{tabular}{lrrrr} 
\noalign{\smallskip}
\hline \hline
\noalign{\smallskip} 
Clump & A  & B & C & Mean \\[2pt] 
\noalign{\smallskip}
\hline
\noalign{\smallskip}
$R$(C$^{18}$O) / pc & 0.9 & 1.1 & 0.7 & 0.9 \\
$\langle R \rangle$(NH$_3$) / pc & 0.07 & 0.13 & 0.09 & 0.10 \\[3pt]
$\Delta v_\mathrm{tot}$(C$^{18}$O$)^{a}$ / km\,s$^{-1}$ & 0.89 & 1.25  & 2.16 & 1.43 \\
$\langle \Delta v_\mathrm{tot} \rangle$(NH$_3$) & 0.57 & 0.46 & 0.72 & 0.58 \\[3pt]
$T_{\mathrm ex}$($^{12}$CO) (K) & 10.2 & 13.1 & 9.2 & 10.8 \\
$\langle T_{k} \rangle$(NH$_3$) / K & 12.9 & 14.6 & 14.2 & 13.9 \\[3pt]
$N_\mathrm{H_2}$(C$^{18}$O)$^{b}$ / 10$^{21}$\,cm$^{-2}$ & 7.1 & 8.4 & 7.8 & 7.8 \\
$N_\mathrm{H_2}$(NH$_3$) / 10$^{21}$\,cm$^{-2}$ & 8.6 & 6.8 & 14.0 & 9.8 \\
$n_\mathrm{H_2}$(NH$_3$)/$n_\mathrm{H_2}$(C$^{18}$O) & 7.7 & 11.6  & 10.0 & 9.8 \\[3pt]
Area(NH$_3$)/Area(C$^{18}$O) & 0.04 & 0.02 & 0.16 & 0.07  \\
$M_\mathrm{cores}/M_\mathrm{clump}$ & 0.04 & 0.02 & 0.19 & 0.08 \\
$V_\mathrm{cores}/V_\mathrm{clump}$ & 0.01 & 0.002 & 0.11 & 0.04 \\[3pt]
\hline
\end{tabular}

\medskip

$^{a}$\,The total line width, $\Delta v_\mathrm{tot}$ of a C$^{18}$O core 
was calculated from the mean line width $\langle \Delta v \rangle$ obtained by 
averaging for each observed position within the half-maximum contour of the 
integrated intensity map and from the dispersion of the mean velocity 
($\delta v_\mathrm{LSR}$): $\Delta v_\mathrm{tot}^{2} = 
\langle \Delta v\rangle^{2} + 8 \ln 2 (\delta \langle v_\mathrm{LSR}\rangle)^{2}$.\\
$^{b}$\,taking into account the revised calibration (Yonekura et al.~1997). 
\end{flushleft}
\end{table}

\begin{figure*}
\centering{
\includegraphics[width=12cm]{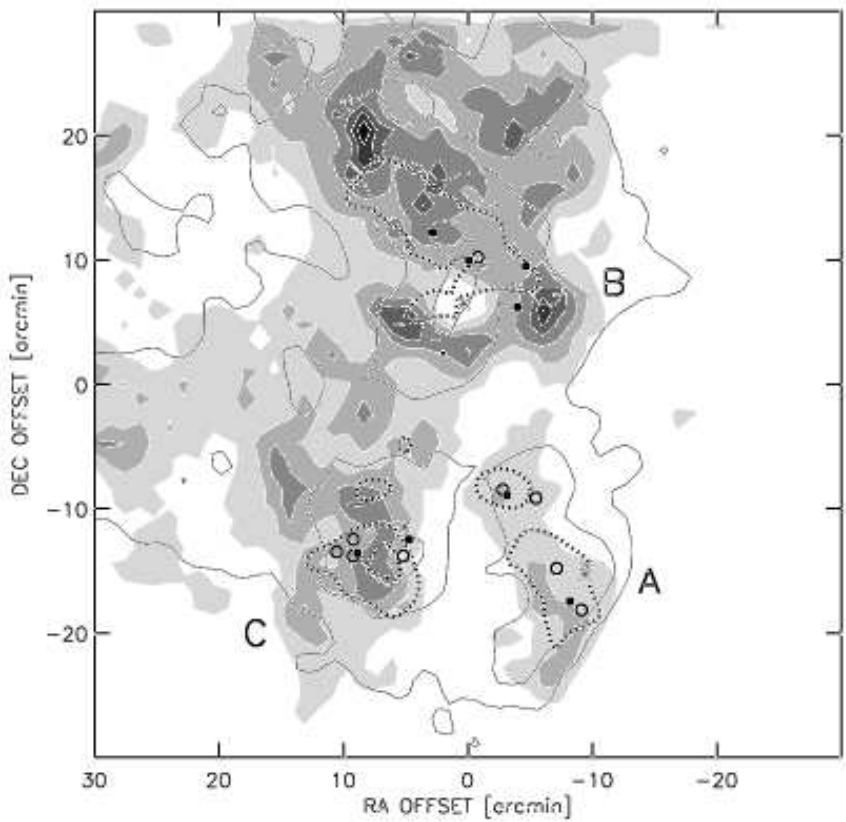}}
\caption{$^{13}$CO (solid contours) and C$^{18}$O  (dotted contours) integrated 
intensity  overlaid on the optical 
extinction map (shading) of L\,1340 constructed from star counts.
Coordinate offsets are given in arcmin with respect to 
RA(2000)=2$^{\rm h}29^{\rm m}42^{\rm s}$ and 
Dec(2000)=+72$^{\rm o}43^{\prime}22^{\prime\prime}$. The lowest 
contour of $^{13}$CO is at 1.0\,K\,kms$^{-1}$, and the increment is 1.5\,K\,kms$^{-1}$. 
The C$^{18}$O contours displayed are 0.45 and 0.75\,K\,kms$^{-1}$. Both the 
lightest shade and the increment is 1\,mag. The $A_\mathrm{V}$ values
displayed are corrected for the foreground extinction. Open circles indicate
the ammonia cores, which probably represent the regions of highest volume densities.
Dots are optically invisible IRAS point sources, and asterisks show the
positions of the RNOs.}
\label{Fig9}
\end{figure*}

Finally, in Fig.~\ref{Fig9} we compare different density cross sections of 
L\,1340, traced by $^{13}$CO, C$^{18}$O, and NH$_3$, with the distribution 
of total cloumn density shown by the visual extinction $A_\mathrm{V}$.
Visual extinction map was constructed from star counts using the USNOFS 
Image and Catalogue Archive\footnote{Operated by the United States Naval 
Observatory, Flagstaff Station (http://www.nofs.navy.mil/data/fchpix/).} 
(see Appendix~B for the details of obtaining $A_\mathrm{V}$).
The angular resolution of $^{13}$CO, C$^{18}$O and $A_\mathrm{V}$ maps  
is equally 3\arcmin.. 
Positions of ammonia cores, embedded YSOs and RNOs are also indicated. 
The amount of the foreground extinction was estimated and subtracted 
from the $A_\mathrm{V}$ values obtained from the star counts
(see Appendix~B). The three clumps can be recognized in the distribution
of $A_\mathrm{V}$, but  some remarkable differences can also be seen 
between the structures shown by the obscuring dust and molecular gas, 
At the southwestern edge of the cloud, in clump~{\it A\/}, similarity of
$^{13}$CO and $A_\mathrm{V}$ suggests that the total amount of 
$A_\mathrm{V}$ originates from the observed molecular gas. The steep 
gradients of both the column density and volume density suggest that 
the gas in this volume has suffered compression from an external shock. 
Both in Clump~{\it B\/} and {
\it C\/} large dark patches can be seen which do not correlate with the molecular 
emission (e.g. around offsets [$14\arcmin,-8\arcmin$], 
[$-4\arcmin,20\arcmin$]). These features indicate diffuse or overlapping 
clumps of high total column density. Together with the compact 
clump~{\it A\/} they give asymmetric, cometary shape to the cloud with a 
`head' pointing towards southwest. Surface distribution of the ammonia cores 
suggests that they have been formed by external compression or magnetic fields 
instead of gravity. Most of them  
({\it A1\/}, {\it A2\/}, {\it B1\/}, {\it C1\/}, {\it C3\/}) are found far 
from the bottom of the gravitational potential well of the embedding clumps, 
indicated either by the peaks of the C$^{18}$O intensity or by the large-scale 
distribution of $A_\mathrm{V}$.

\subsection{Comparison with other clouds}
\label{Sect_5.3}

Comparison of properties of ammonia cores in L\,1340 with JMA's data base
(their Tables B9--B20) shows that the typical sizes, kinetic temperatures, 
line widths and masses of ammonia cores are increasing in the order of
Taurus $\rightarrow$ Ophiuchus $\rightarrow$ Perseus $\rightarrow$ L\,1340 
$\rightarrow$ Orion~B $\rightarrow$ Orion~A. The IRAS luminosities 
do not show this trend, being lower in L\,1340 than in Perseus. 
A reason for this departure from the trend may be the difference in cloud 
distances. L\,1340 is the most distant among the clouds listed above,
therefore a considerable fraction of YSOs born in it might remained undetected 
by IRAS.  We have shown in
Sect.~\ref{Sect_5.2} that the slope $\alpha$ of the line width--size 
relation also shows the Taurus $\rightarrow$ L\,1340 
$\rightarrow$ Orion~B trend, suggesting that properties of cores and newborn 
stars are related to large-scale interstellar processes. Comparison of 
observational results with the continuously improving numerical 
simulations of such processes will lead to a better understanding of the 
cloud formation and evolution. This is, however, beyond the scope of 
the present paper.

\section{Summary of the results} 
\label{Sect_6} 
 
The main results of ammonia observations of L\,1340 are summarized as follows.
\begin{itemize}
\item[(i)] Mapping of the whole area of the cloud where C$^{18}$O 
emission indicated high gas volume density ($n \geq 10^3\,\mathrm{cm}^{-3}$)
in the NH$_3$(1,1) and (2,2) lines resulted in the detection of 10 dense cores 
in L\,1340. We found the embedding cores of six candidate YSOs, namely
IRAS 02238+7222, 02249+7230, F02256+7249, 02267+7226, 02276+7225 and
F02277+7226. An additional object, IRAS 02263+7251 is associated with 
weak ammonia emission near the detection limit.
\item[(ii)] $M > M_\mathrm{BE}$ for most cores, indicating that they 
are able to form stars.
\item[(iii)] The cores with and without embedded YSOs differ from each other 
in nonthermal line width. The nonthermal velocity dispersion is subsonic in the 
starless group. Outflows from the known embedded protostars cannot account for 
the high $\Delta v_\mathrm{NT}$ values of cores associated with IRAS
sources. Therefore the two groups differ from each other not only
in evolutionary state. The nonthermal line width of cores is probably related 
to the masses of stars being formed in them.
\item[(iv)] The velocity structure of L\,1340\,C revealed two clumps moving
with supersonic velocities with respect to each other. Star formation  
has possibly been triggered by clump collision  in this region.
\item[(v)] The relations between the physical properties of NH$_3$ cores 
and the $^{13}$CO and C$^{18}$O clumps are consistent with the scenario of 
turbulent fragmentation. 
\item[(vi)] The nonthermal line width--size relation revealed by  NH$_3$,
 C$^{18}$O, $^{13}$CO, and \ion{H}{i} shows self-similar structure between
0.1--40\,pc. Its slope is $\alpha=0.41\pm0.06$.
\item[(vii)] The typical size, kinetic temperature and line width of 
NH$_3$ cores rank L\,1340 between the high mass star forming regions 
Perseus and Orion~B, whereas its total mass is some two orders of magnitude 
smaller, and no high mass stars have been formed in it. Apparently most of 
the ISM around L\,1340 is in the form of \ion{H}{i}.
\end{itemize}

\begin{acknowledgements}
This research was supported by the Hungarian OTKA grants T022946, T024027, 
T034998, T034584, and T037508. We also received support from the 
German--Hungarian Technological and Scientific Cooperation Project
No. 121. L. V. T. acknowledges financial support from the Academy of Finland
grant No. 174854. This research has made use of the USNOFS Image and Catalogue 
Archive operated by the United States Naval Observatory, Flagstaff Station
(http://www.nofs.navy.mil/data/fchpix/).
We thank to P\'eter \'Abrah\'am and Attila Mo\'or for their help in
handling the \ion{H}{i} data, and L\'aszl\'o Szabados for careful 
reading of the manuscript.
\end{acknowledgements}

\appendix

\section{Distribution of the H{\footnotesize I} around L\,1340}

\begin{figure}
\resizebox{\hsize}{!}{\includegraphics{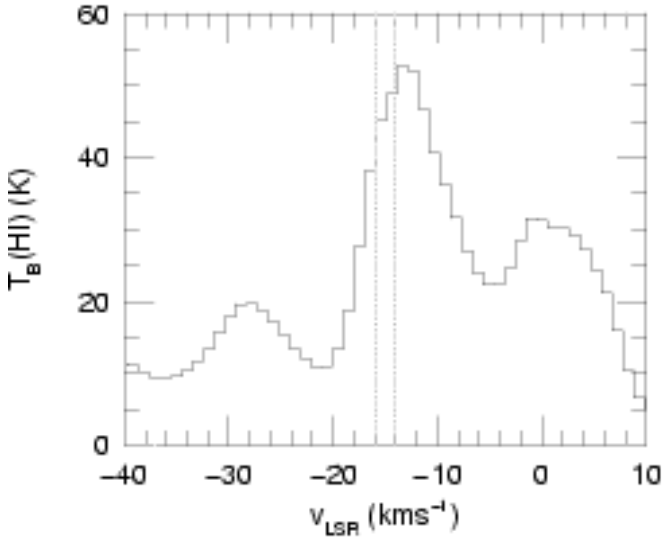}}
\caption{\ion{H}{i} spectrum at $l=130\fdg5, b=+11\fdg5$  taken from the
Leiden--Dwingeloo survey. The angular resolution of the survey is 0\fdg5,
and the spectral resolution is 1.03\,kms$^{-1}$. Dotted vertical lines indicate the 
radial velocity range of the molecular cloud.}
\label{FigA1}
\end{figure}

The angular resolution of the Leiden--Dwingeloo \ion{H}{i} survey 
data is 0\fdg6, corresponding to 6.3\,pc at the distance of L\,1340, 
and the velocity  resolution is 1.03\,km\,s$^{-1}$.
Fig.~\ref{FigA1} shows the \ion{H}{i} spectrum at (130\fdg5,+11\fdg5),
and in the velocity interval 
$-40\,\mathrm{km\,s}^{-1} < v_\mathrm{LSR} < +10\,\mathrm{km\,s}^{-1}$. 
The peak at 0\,kms$^{-1}$ probably corresponds to the atomic 
cloud associated with the molecular cloud L\,1333 located
at a distance of 180\,pc around (l,b)=(128\fdg9,13\fdg7), and at the same 
mean radial velocity (Obayashi et al.~\cite{OKSYF}). The highest peak
of the spectrum at $v_\mathrm{LSR}=-13$\,kms$^{-1}$ represents 
the \ion{H}{i} cloud enveloping L\,1340. The characteristic line width 
of this spectral feature is $\sim$\,7\,km\,s$^{-1}$.

\begin{figure}
\resizebox{\hsize}{!}{\includegraphics{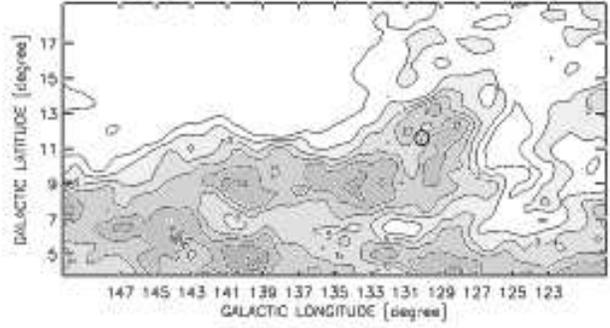}}
\caption{Distribution of the neutral hydrogen over a $30\degr \times 15\degr$
area including L\,1340, integrated over the velocity interval 
$-18\,\mathrm{kms}^{-1} < v_\mathrm{LSR} < -8\,\mathrm{kms}^{-1}$, taken from 
the Leiden--Dwingeloo survey. The only known molecular cloud in this area is
L\,1340, marked by the circle. Column densities were derived  assuming optically 
thin emission. The lowest contour is at $3\times\,10^{20}$\,cm$^{-2}$,
and the increment is 10$^{20}$\,cm$^{-2}$.}
\label{FigA2}
\end{figure}

Fig.~\ref{FigA2} shows the distribution of the neutral 
hydrogen integrated over the velocity interval 
$-18\,\mathrm{km\,s}^{-1} < v_\mathrm{LSR} < -8\,\mathrm{km\,s}^{-1}$.
 A large, elongated  \ion{H}{i} structure can be seen in the area 
$126\degr \leq l \leq 142\degr$ and $+7\degr \leq b \leq +13\degr$ in this 
radial velocity interval. Its radius, derived from the area within the 
half-maximum contour, is  $\sim$\,38\,pc. The apparent 
local minimum in \ion{H}{i} 
near the molecular cloud may result both from 
self-absorption and conversion of a part of hydrogen into molecules. 
The \ion{H}{i} column densities displayed 
in Fig.~\ref{FigA2} were estimated assuming optically 
thin emission, i.e. using the relationship 
\begin{equation}
N(\ion{H}{i})=1.8224\times10^{18} \int \!T_b dv \mathrm{cm^{-2}(K\,kms^{-1})^{-1}}
\label{eq_A1} \end{equation}
(Rohlfs \& Wilson~2000). The mass of the \ion{H}{i} structure,
estimated via summing up the column densities inside the half-maximum 
contour is $M$(\ion{H}{i})$ \geq 2\,\times 10^{4}$\,M$_{\sun}$, comparable with 
those of some well known nearby molecular cloud complexes (e.g. Taurus),
and an order of magnitude larger than the mass of the molecular cloud.	

\section{Deriving the distribution of $\mathbf{A_V}$ from star counts}

We used the USNOFS Image and Catalogue Archive to derive the distribution of the visual extinction
$A_\mathrm{V}$ in a field of $1\degr \times 1\degr$ containing  L\,1340. 
We counted the stars on overlapping circles of 3$\arcmin$ in diameter, 
the centres of which were distributed on a regular grid with step of 30$\arcsec$.
We removed from the stellar list  all known candidate pre-main sequence stars 
associated with the cloud, and all identified foreground stars.
We derived $A_\mathrm{V}$ from $R$ star counts using the method
described by Dickman~(\cite{Dickman}). The extinction-free reference area
was a field 20$\arcmin\times\,20\arcmin$ centred on RA(2000)=1$^\mathrm{h}52^\mathrm{m}$, 
Dec(2000)=+73\degr15\arcmin. $A_\mathrm{V}$ values obtained in 
this manner saturate at 6~mag. 

We estimated the contribution of the foreground diffuse matter to $A_\mathrm{V}$ 
with the aid of the neutral hydrogen spectra (see Fig.~\ref{FigA1}). 
Assuming optically thin radiation we  used the relationship 
\ref{eq_A1} to derive hydrogen column densities 
from the spectra, and regarded as foreground all the \ion{H}{i} gas at velocities 
$v_\mathrm{LSR} > -6$\,kms$^{-1}$, as well as half of the gas  
at $-$20\,kms$^{-1} < v_\mathrm{LSR} < -$6\,kms$^{-1}$.
Taking the average of four \ion{H}{i} spectra covering the face of 
L\,1340, and using the relationship 
$N$(\ion{H}{i}) $\approx 2\times10^{21}$\,[cm$^{-2}$/mag] A$_\mathrm{V}$
(Spitzer~\cite{Spitzer}) we obtained $A_{\mathrm{V},f} \approx$ 0.55 mag for the foreground 
extinction to be subtracted from the $A_\mathrm{V}$ values derived from the 
star counts.

\end{document}